\documentclass[conference]{IEEEtran}
\def\IEEEtitletopspace{0.3in}
\IEEEoverridecommandlockouts
\usepackage{amsmath,amssymb,amsfonts,amsthm}
\usepackage{graphicx}
\usepackage{enumerate}

\newcommand{\Par}[1]{\left(#1\right)}
\newcommand{\Bkt}[1]{\left[#1\right]}
\newcommand{\Norm}[1]{\left\|#1\right\|}
\newcommand{\Set}[1]{\left\{#1\right\}}

\newcommand{\Reals}{{\mathbb{R}}}
\newcommand{\PInts}{{\mathbb{Z}_{\geq 0}}}
\newcommand{\Sp}[1]{{\Reals}^{d_{#1}}}

\newcommand{\Seq}{{{S}}} 
\newcommand{\SeqSp}[1]{{\Seq}{\Par{#1}}} 

\newcommand{\eps}{{\varepsilon}}
\newcommand{\Flow}{{\varphi}}   
\newcommand{\dflow}{{\rho}}     
\newcommand{\Nets}{{\mathfrak{N}}}

\newcommand{\dd}{{\mathrm{d}}}

\newcommand{\Indic}{{\mathbf{1}}}

\newcommand{\States}{{X}}
\newcommand{\Controls}{{\mathbb{U}}}
\newcommand{\Param}{{\omega}}
\newcommand{\Tau}{{\mathfrak{t}}}

\newcommand{\Hyp}{{\mathcal{H}}}
\newcommand{\Id}{{\mathrm{id}}}

\newtheorem{theorem}{Theorem}
\newtheorem*{theorem*}{Theorem}
\newtheorem{lemma}{Lemma}
\newtheorem{sublemma}{Sublemma}

\theoremstyle{definition}
\newtheorem{definition}{Definition}

\def\BibTeX{{\rm B\kern-.05em{\sc i\kern-.025em b}\kern-.08em
    T\kern-.1667em\lower.7ex\hbox{E}\kern-.125emX}}

\begin{document}

\title{Universal approximation of flows of control systems by recurrent neural networks
\thanks{
    $^{1}$Digital Futures and Division of Decision and Control Systems,
    KTH Royal Institute of Technology,
    SE-100 44 Stockholm, Sweden.
    Email: {\tt\small \{aguiar, kallej\}@kth.se},
    $^{2}$Control Systems Group, EE Dept., Eindhoven University of Technology,
    P.O. Box 513, 5600 MB Eindhoven, The Netherlands. Email:
    {\tt\small am.das@tue.nl}
}%
}
\author{
\IEEEauthorblockN{Miguel Aguiar$^1$} \and 
\IEEEauthorblockN{Amritam Das$^2$} \and
\IEEEauthorblockN{Karl H. Johansson$^1$}
}

\maketitle

\begin{abstract}
We consider the problem of approximating flow functions of continuous-time dynamical systems with inputs.
It is well-known that continuous-time recurrent neural networks are universal approximators of this type of system.
In this paper, we prove that an architecture based on discrete-time recurrent
neural networks universally approximates flows of continuous-time 
dynamical systems with inputs.
The required assumptions are shown to hold for systems whose
dynamics are well-behaved ordinary differential equations and
with practically relevant classes of input signals.
This enables the use of off-the-shelf solutions for learning such flow functions in continuous-time from sampled trajectory data.
\end{abstract}


\vspace{-1em}

\section{Introduction}



The advantage of continuous-time models for learning dynamics
has been pointed out in a number of recent
works~\cite{ChenEtAl18, DeBrouwerEtAl19, BeintemaEtAl22}.
Such models naturally handle irregularly sampled or missing data,
and are the natural model class for most physical systems.

Some approaches for continuous-time identification have been
proposed~\cite{GarnierYoung14}, but for nonlinear systems the
majority of research concentrates on discrete-time models~\cite{SchoukensLjung19}.
A number of modeling approaches have arisen using ideas and model classes
from classical and deep machine learning.
It is well known that continuous-time recurrent neural networks can approximate
large classes of continuous-time dynamical systems with inputs,
see Sontag~\cite{Sontag92}, Li~et~al~\cite{LiEtAl05}
and references therein.
In fact, these networks are able to approximate flows of
stable continuous-time systems over unbounded
time intervals~\cite{HansonRaginsky20},
as well as more general input-output operators~\cite{HansonEtAl21}.
Neural Ordinary Differential Equations~(Neural ODEs)~\cite{ChenEtAl18}
are a particular class of continuous-time models proposed
to replace standard network layers appearing in models used
for common learning tasks, and have been shown to be
competitive with state-of-the-art models in system
identification~\cite{RahmanEtAl22}.
In~\cite{ForgionePiga21} some specific architectures and
learning methods for identifying differential equation models
of control systems using neural networks are presented.

An assortment of related methods have been proposed for modeling autonomous
systems with applications in the 
physical sciences~\cite{GenevaZabaras22, FloryanGraham22, BruntonEtAl16}.
For methods based on Koopman~operator approximation in particular, extensions
to certain classes of systems with inputs are possible~\cite{BevandaEtAl21}.
Physics-informed learning has also emerged as a paradigm for learning
solutions of ordinary and partial differential equations from 
data~\cite{RaissiEtAl19,KarniadakisEtAl21}.
These methods incorporate a set of differential equations
known to be satisfied by the data as a regulariser in the
loss function used to train the network,
and can also be used to identify parameters in the equations.

In contrast to the majority of these approaches,
the class of methods known as neural operator methods
attempt to directly learn the solution operator of a differential
equation, that is, the operator mapping initial conditions, forcing terms
and parameters to the corresponding solution, rather than identifying 
the governing equations~\cite{KissasEtAl22, BilosEtAl21,LiEtAl21,LuEtAl21}.
The focus is then on engineering architectures with the appropriate
inductive biases for a particular class of problems.

In this paper, we consider the problem of approximating the flow function
of a dynamical system, that is, the solution operator mapping
initial conditions and control inputs to the corresponding
trajectory of the system.
Exploiting the discrete structure of commonly used classes
of control inputs, we show that the flow function can be exactly
represented by a discrete-time dynamical system.
This motivates the use of (discrete-time) Recurrent Neural Network (RNN)
architectures to learn flow functions from data.
We propose one such architecture which ensures that the trajectories of
the learned model are continuous.
In previous work~\cite{AguiarEtAl23}, we have shown through
numerical experiments that the architecture successfully learns
flows of oscillators with complex dynamics, and have investigated
its generalisation performance.

In this paper, our contribution is twofold.
Firstly, we prove that the proposed architecture is a universal
approximator for flow functions of control systems,
which guarantees the well-posedness of the learning problem
that we formulate mathematically in~\cite{AguiarEtAl23}.
Secondly, we show by system-theoretic arguments
that the required assumptions hold for
systems whose dynamics are given by well-behaved
ODEs, with rather general and
practically relevant classes of input signals.

This approach has a number of advantages in comparison
with methods based on learning the right-hand side of
a differential equation.
Errors in the learned dynamics can be propagated and
affect long-term prediction performance.
When the flow is directly approximated, the need for
integration is obviated.
This has the additional advantage of reducing the computational burden
at both training and prediction time.
In effect, under our formulation, the problem of learning
a flow function amounts to a standard regression problem,
and thus enables the use of off-the-shelf learning frameworks
for training the model.
At prediction time, one can query the solution map at any time
instant, and, since the model uses standard neural network
components, gradients of the flow with respect to, e.g.,
initial conditions or control values can be computed
in a straightforward manner through automatic differentiation.
Furthermore, the approach is able to accommodate more general
classes of systems than those with dynamics given by ODEs.

The rest of the paper is organised as follows.
Section~\ref{sec:prelim} introduces notation and some basic definitions.
In Section~\ref{sec:problem-formulation} we describe the proposed architecture and the considered class of input signals.
This is followed by the statement of Theorem~\ref{thm:ua-flows}
in Section~\ref{sec:statement}.
In Section~\ref{sec:ua-dt} we give a proof that discrete-time RNNs
are universal approximators, an essential step in the
proof of Theorem~\ref{thm:ua-flows} given in Section~\ref{sec:ua-flows}.
Section~\ref{sec:ua-odes} treats the case of flows of ODEs and the 
assumptions of Theorem~\ref{thm:ua-flows}
are shown to hold in that setting.
A numerical example is briefly discussed in 
Section~\ref{sec:num-example},
and concluding remarks are given in Section~\ref{sec:conclusion}.

\vspace{-1em}
\section{Preliminaries}\label{sec:prelim}
\subsection{Notation}
\vspace{-0.7em}
The indicator function of a set $A$ is written $\Indic_{A}$,
and the identity function on $A$ is written $\Id_A$.
Sequences are written $(z_k)_{k=0}^{\infty}$, or in short-hand $(z_k)$.
The space of sequences with values in $A$ is written
${\SeqSp{A} := \Set{(z_k)_{k=0}^{\infty}: z_k \in A}}$.
The space of continuous functions $f : A \to \Reals^n$ on a compact set
${A \subset \Reals^m}$ is written $C_n(A)$.
For $A \subset \Reals^n$ and $\eps > 0$, $N_\eps(A)$ denotes the (closed) $\eps$-neighbourhood of $A$,
i.e., the set of points at most $\eps$ distance away from $A$.
If $A$ is compact, then so is $N_\eps(A)$.
Vectors $v \in \Reals^{d}$ are written $v = (v_1, \dots, v_d)$.
We denote by $\Norm{\cdot}$ the Euclidean norm on $\Reals^d$,
and for a matrix $M \in \Reals^{m \times n}$,
$\Norm{M}$ denotes the induced operator norm.

\vspace{-0.5em}
\subsection{Flows of controls systems in continuous-time}\label{ssec:flows-defn}
In this paper we consider finite-dimensional
time-invariant control systems in continuous-time with state evolving in
an open set $\States \subset \Sp{x}$.
Such systems can be described abstractly by a \textit{flow function}
\begin{equation}
    {\Flow: \Reals_{\geq 0} \times \States \times \Controls \to \States}
\end{equation}
where $\Controls$ is a given set of (control) inputs
$u: \Reals_{\geq 0}\to \Reals^{d_u}$.
The flow satisfies the following properties (cf.~Sontag~\cite[Chapter~2]{Sontag98}):
\begin{itemize}
    \item Identity: $\Flow(0, x, u) = x$
    \item Semigroup: $\Flow(s + t, x, u) = \Flow(t, \Flow(s, x, u), u^s)$
\end{itemize}
for all $x \in X$, $u \in \Controls$ and $s, t \geq 0$.
Here $u^s \in \Controls$ denotes the input $u$ shifted by $s > 0$ time units, i.e.,
$u^s(t) := u(t + s)$.
The function $t \mapsto \Flow(t, x, u), t \geq 0$ is the trajectory of the system
with initial state $x$ when the applied control is $u$.

\vspace{-0.5em}
\subsection{Neural networks as function approximators}
In this paper, a \emph{(feedforward) neural network} is any function $h: \Reals^{m} \to \Reals^n$
which can be written as
\begin{equation}
\label{eq:ff-network}
    h(x) = C \sigma_p(Ax + b) + d, \ x \in \Reals^m
\end{equation}
for $A \in \Reals^{p \times m}$, $b \in \Reals^p$,
$C \in \Reals^{n \times p}$, $d \in \Reals^{n}$.
Here ${\sigma_p: \Reals^p \to \Reals^p}$ is a diagonal mapping 
such that the \emph{activation function} $\sigma: \Reals \to \Reals$
is applied to each coordinate, i.e.,
${\sigma_p(v) = (\sigma(v_1), \dots, \sigma(v_p))}$ 
for $v \in \Reals^p$. In practical applications, these are usually known as networks with
\emph{one hidden layer}.

We let $\Nets^{m, n}_{\sigma, p}$ be the class of such networks and define
    $\Nets^{m, n}_{\sigma} :=
        \cup_{p = 1}^{\infty} \Nets^{m,n}_{\sigma, p}.$
Throughout the paper, we shall assume that 
${\sigma : \Reals \to \Reals}$ is a fixed bounded,
continuous and nonconstant function.
Under this assumption it is well-known~\cite{Hornik91} that
$\Nets^{n, m}_\sigma$ is dense in $C_n(K)$ for any compact set
$K \subset \Reals^m$.
That is, for any continuous function $f: K \to \Reals^n$ and $\eps > 0$
there is a network $h \in \Nets^{n, m}_{\sigma}$ such that
    $\sup_{x \in K}\Norm{f(x) - h(x)} < \eps$.

Let $\Nets^{0}_{\sigma, p} \subset \cup_{q \geq p} \Nets^{q, p}_{\sigma, p}$ be the class of feedforward networks for which $C = I$ and $d = 0$ in~\eqref{eq:ff-network}. A \emph{Recurrent Neural Network} (RNN) is then simply a difference equation whose right-hand side is a network in $\Nets^0_{\sigma, p}$ for some $p \geq 0$:
\begin{definition}[RNN]
An RNN is a difference equation of the form
\begin{equation*}
    z_{k + 1} = \sigma_{d_z}(A z_k + B u_{k} + b), \ k \in \PInts,
\end{equation*}
where $z \in \Sp{z}$, $u \in \Sp{u}$.
\end{definition}

\vspace{-1em}
\section{Architecture Definition}\label{sec:problem-formulation}
In this section we define
a discrete-time RNN-based architecture to approximate flow functions of continuous-time dynamical systems.
We focus in particular on systems for which the trajectories $t \mapsto \Flow(t, x, u)$
are continuous in time $t$.
This is the case when $\Flow$ arises from a differential equation, differential algebraic equation,
but excludes e.g. hybrid systems with state jumps.
We shall show that $\Flow$ can be approximated by a function $\hat\Flow$
on a finite time interval, where $\hat\Flow(t, x, u)$ is computed by an RNN.
In the following sections we make this precise.

\subsection{Class of inputs}\label{ssec:input-class}
In order to approximate $\Flow$, we must impose some structure on $\Controls$.
In practice, the majority of systems are controlled by
a computer with a zero-order hold digital-to-analog converter,
so that the input signal will be piecewise constant,
with the control value changing at regular time instants
with some period $\Delta > 0$.
Occasionally, first- or higher-order polynomial parameterisations are also used.
In this paper we consider a general parameterisation of control inputs
which encompasses all of these cases.
Namely, we assume that the control can be parameterised by a sequence of finite-dimensional
parameters $(\Param_k)_{k=0}^{\infty} \subset \Sp{\Param}$ as follows:
\begin{equation}\label{eq:control-param}
    u(t) = \sum_{k=0}^\infty{
        \alpha{\Par{\Param_{k}, \frac{t}{\Delta}}} 
            \Indic_{[k\Delta, (k + 1)\Delta)}(t)
    }, \ t \geq 0.
\end{equation}
Here $\alpha: \Sp{\Param} \times \Reals_{\geq 0} \to \Sp{u}$
is periodic with period~$1$ in its second argument.
In other words, we have for each $k \geq 0$
\begin{equation*}
    u(t) = \alpha(\Param_k, t/\Delta), \ k\Delta \leq t < (k+1)\Delta.
\end{equation*}

The simplest example is given by $\alpha(\Param, t) := \Param$,
corresponding to the case of piecewise constant controls with
period~$\Delta$.

Throughout the paper we assume that $\Delta$ and the function $\alpha$
are fixed and known.
For a set $\Omega \subset \Sp{\Param}$ we define the set 
$\Controls(\Omega)$ of controls $u$ parameterised by
sequences in $\SeqSp{\Omega}$ according to~\eqref{eq:control-param}, i.e.,
\begin{equation}
\begin{aligned}\label{eq:control-set}
    \Controls(\Omega) := \bigg\{
        u &: \Reals_{\geq 0} \to \Sp{u} : \Par{\Param_k}_{k=0}^{\infty} \in S(\Omega),\\
        &\quad u(t) = \sum_{k=0}^\infty{
            \alpha{\Par{\Param_{k}, \frac{t}{\Delta}}} 
                \Indic_{[k\Delta, (k + 1)\Delta)}(t)
        } \bigg\}.
\end{aligned}
\end{equation}


\subsection{Representing flows by discrete-time systems}
We let $u_\Param$ be the control generated by the constant sequence 
with value $\Param$, so that
    $u_\Param(t) = \alpha(\Param, t/\Delta), \ t \geq 0$
and define the function $\Phi: [0, 1] \times X \times \Sp{\Param} \to X$ by
\begin{equation*}
    \Phi(\tau, x, \Param) := \Flow(\tau\Delta, x, u_\Param).
\end{equation*}
Fix $t \in \Reals_{\geq 0}$ and define
${k_t := \lfloor t/\Delta \rfloor}, \
{\tau_t := (t - k_t\Delta)/\Delta}$.
The value of $\Flow(t, x, u)$ can be computed recursively by $\Phi$ as follows:
\begin{equation}\label{eq:flow-recursive}
\begin{aligned}
    x_0 &= x \\
    x_{k + 1} &= \Phi(1, x_k, \Param_{k}), \ 0 \leq k < k_t \\
    x_{k_t + 1} &= \Phi(\tau_t, x_{k_t}, \Param_{k_t}) = \Flow(t, x, u).
\end{aligned}
\end{equation}
This can be seen as representing $\Flow$ by a discrete-time system
with inputs $(\tau, \Param) \in [0, 1] \times \Sp{\Param}$.
Note that such a representation does not amount to a \emph{discretisation} of~$\Flow$, so that no loss of information or generality is incurred,
and we are able to compute the flow~$\Flow$
at any instant of time through this correspondence.

The discrete-time system defined by~\eqref{eq:flow-recursive} can be approximated by an RNN as follows.
Let $x \in X$, $t \geq 0$ and $u \in \Controls$ be parameterised according to~\eqref{eq:control-param} by a sequence $(\Param_k)$.
Fixing networks $h \in \Nets^0_{\sigma, d_z}$, $\beta \in \Nets^{d_x, d_z}_{\sigma}$
and $\gamma \in \Nets^{d_z, d_x}_{\sigma}$, compute the sequence
\begin{equation}\label{eq:hidden-states}
\begin{aligned}
    z_0 &= \beta(x) \\
    z_{k + 1} &= h(1, z_k, \Param_{k}), \ 0 \leq k < k_t \\
    z_{k_t + 1} &= h(\tau_t, z_{k_t}, \Param_{k_t})
\end{aligned}
\end{equation}
and set
\begin{equation*}
    \hat\Flow(t, x, u) = \gamma((1 - \tau_t)z_{k_t} + \tau_t z_{k_t + 1}).
\end{equation*}
The interpolation guarantees that $\hat\Flow$ is continuous in $t$.
Note that it does not amount to a linear interpolation, 
as $z_{k_t + 1}$ depends on $\tau_t$.

In order to express $\hat\Flow$ explicitly, the following definition is useful,
and will be used throughout the following sections.
\begin{definition}[Recursion Map]
Let $f: A \times B \to A$.
The associated \emph{recursion map} $\dflow_f : \PInts \times A \times S(B) \to A$
is defined as
\begin{equation}\label{eq:recursion-map}
\begin{aligned}
    \dflow_f(0, x, (u_k)) &= x, \\
    \dflow_f(n + 1, x, (u_k)) &=
            f(\dflow_f(n, x, (u_k)), u_{n}), \ n \geq 0.
\end{aligned}
\end{equation}
\end{definition}

Now let $(\Tau^t_k)_{k = 0}^{\infty} \in \SeqSp{[0, 1]}$ be defined by
\begin{equation}\label{eq:tau-seq}
    \Tau^t_k = \begin{cases}
        1, & 0 \leq k < k_t \\
        \tau_t, & k = k_t \\
        0, & k > k_t.
    \end{cases}
\end{equation}
Then we can rewrite~\eqref{eq:hidden-states} as\footnote{%
With some abuse of notation, we interpret $h$ as a function mapping ${\Sp{z}\times([0, 1]\times\Sp{\Param})}$ to $\Sp{z}$.
}
${z_k = \dflow_h(k, \beta(x), (\Tau^t_k, \Param_k))}$, $k \geq 0$,
and thus $\hat\Flow$ can be written
\begin{equation*}
\begin{aligned}
    \hat\Flow(t, x, u) = \gamma[&(1 - \tau_t)
        \dflow_h(k_t, \beta(x), (\Tau^t_k, \Param_k)) \\
        &~{}+ \tau_t \dflow_h(k_t + 1, \beta(x), (\Tau^t_k, \Param_k))].
\end{aligned}
\end{equation*}
We let $\Hyp$ denote the set of functions
${\hat\Flow: \Reals_{\geq 0} \times \States \times \Controls \to \Sp{x}}$
defined in this way, that is,
\begin{equation}\label{eq:hyp}
\begin{aligned}
    \Hyp := \bigg\{
        &{\hat\Flow: \Reals_{\geq 0} \times \States \times \Controls \to \Sp{x}} :
        d_z \in \PInts, \\ 
        &\gamma \in \Nets^{d_z, d_x}_{\sigma}, \ h \in \Nets^0_{\sigma, d_z},
            \ \beta \in \Nets^{d_x, d_z}_{\sigma}, \\
        &\hat\Flow(t, x, u) = \gamma[(1 - \tau_t)
            \dflow_h(k_t, \beta(x), (\Tau^t_k, \Param_k)) \\
        &\quad\quad\quad\quad\quad\,~{}+ 
            \tau_t \dflow_h(k_t + 1, \beta(x), (\Tau^t_k, \Param_k))]
    \bigg\}.
\end{aligned}
\end{equation}
For a more detailed explanation and motivation of the architecture just described,
the reader is referred to~\cite{AguiarEtAl23}.

\section{Statement of the Main Result}\label{sec:statement}
\begin{theorem}\label{thm:ua-flows}
Suppose the flow of a control system ${\Flow: \Reals_{\geq 0} \times \States \times \Controls \to \States}$ satisfies the following conditions:
\begin{enumerate}
    \item Given a compact set $K_{\Param} \subset \Sp{\Param}$, define $\Controls(K_{\Param})$ according to \eqref{eq:control-set}. Then $\Controls(K_\Param) \subset \Controls$, i.e., for any $u \in \Controls(K_{\Param})$,
    the corresponding trajectory $\Flow(\cdot, x, u)$
    is well-defined for all ${x \in \States}$. 
    \item   The function $\Phi: [0, 1] \times X \times \Reals^{d_\omega} \to X$ defined as
    $$ \Phi(\tau, x, \omega) := \Flow(\tau\Delta, x, u_\omega), \ u_\omega(t) = \alpha(\omega, t/\Delta)$$
    is right-differentiable at ${\tau = 0}$ for every 
    ${(x, \Param) \in X \times \Sp{\Param}}$.
    \item   The function ${\Psi: [0, 1] \times \States \times \Sp{\Param}}$ defined as
    \begin{equation}\label{eq:psi-def}
        \Psi(\tau, x, \omega) := \begin{cases}
            x + \tau^{-1}(\Phi(\tau, x, \omega) - x), \ \tau \in (0, 1] \\
            \displaystyle\lim_{t \downarrow 0} \Bkt{x + t^{-1}(\Phi(t, x, \omega) - x)}, \ \tau = 0 
        \end{cases}
    \end{equation}
    is continuous and locally Lipschitz in $x$.
\end{enumerate}
Then, for any $\eps > 0$, $T \geq 0$ and compact sets ${K_x \subset X}$, ${K_\Param \subset \Sp{\Param}}$, there exists $\hat\Flow \in \mathcal{H}$, defined according to \eqref{eq:hyp}, such that
        $\Norm{ \Flow(t, x, u) - \hat\Flow(t, x, u)} < \eps $
    holds for all $t \in [0, T]$, $x \in K_x$ and $u \in \Controls(K_{\Param})$.
    Furthermore, $\gamma$ and $\beta$ in~\eqref{eq:hyp}
    can be chosen to be affine with $\gamma\circ\beta = \Id_{\Sp{x}}$.
\end{theorem}

Note that assumptions~2 and~3 implicitly represent assumptions on $\Flow$ and $\alpha$.
In Section~\ref{sec:ua-odes} we shall give conditions under which they are satisfied for flows of differential equations.

\section{Universal Approximation of Discrete-Time Systems}\label{sec:ua-dt}

In this section we give a proof that discrete-time
RNNs are universal approximators of discrete-time systems,
a fact that will be used in the proof of Theorem~\ref{thm:ua-flows}.

\begin{theorem}[Universal approximation for discrete-time dynamical systems]\label{thm:ua-discrete}
    Let $f : \Sp{x} \times \Sp{u} \to \Sp{x}$ be a continuous function that
    is locally Lipschitz in the first variable, in the sense that for any
    compact set ${K \subset \Sp{x}}$ there exists a locally bounded function
    ${\nu_K: \Sp{u} \to \Reals_{\geq 0}}$ such that
    \begin{equation*}
        \Norm{f(x_2, u) - f(x_1, u)} \leq \nu_K(u)\Norm{x_2 - x_1}, \ x_1, x_2 \in K.
    \end{equation*}
    Then for any $\eps > 0$, $N \in \PInts$ and compact sets
    $K_x \subset \Sp{x}$ and $K_u \subset \Sp{u}$
    there exist networks $h \in \Nets^0_{\sigma, d_z}$,
    $\gamma \in \Nets_\sigma^{d_z, d_x}$ and
    $\beta \in \Nets_\sigma^{d_x, d_z}$
    such that for any $x \in K_x$ and $u \in \SeqSp{K_u}$ we have
    \begin{equation}\label{eq:sim}
        \Norm{\dflow_f(n, x, u) - \gamma(\dflow_h(n, \beta(x), u))} < \eps, \ n = 0, \dots, N,
    \end{equation}
    where $\dflow_f$ is defined as in~\eqref{eq:recursion-map}.
    Furthermore, $\gamma$ and $\beta$ can be chosen to be affine with $\gamma\circ\beta = \Id_{\Sp{x}}$.
\end{theorem}

Noe that $h$ above defines an RNN. Therefore, Theorem 
 \ref{thm:ua-discrete} states
that any discrete-time dynamical system with a locally Lipschitz right-hand side
can be approximated in the sense of~\eqref{eq:sim} by an RNN.
This is a well-known fact~\cite{Sontag92,SchaferZimmermann06},
but for reference we include a full proof under the stated assumptions,
as this result is an important step in the proof of Theorem~\ref{thm:ua-flows},
and there we will in particular use the fact that $\gamma \circ \beta = \Id_{\Sp{x}}$
and that these maps are affine.

\begin{proof}

The case $N = 0$ is trivial and $N = 1$ corresponds to
the standard universal approximation theorem proved in Hornik~\cite{Hornik91},
so we assume $N \geq 2$ in what follows.

Define the sets $K^0_x, \dots, K^{N - 1}_x$
recursively by ${K^{n + 1}_x = {f(K^n_x, K_u)}}$ with $K^0_x = K_x$.
By continuity of $f$, the $K^n_x$ are compact.
For any input sequence $u \in \SeqSp{K_u}$ and initial state $x_0 \in K_x$, we then have that
\begin{equation*}
    \dflow_f(n, x_0, u) \in K^n_x, \ n = 0, 1, \dots, N - 1.
\end{equation*}

Define also $L_f^n, \eta_f^n \geq 0$ and sets $\tilde{K}^n$,
$n = 1, \dots, N-1$ recursively as follows:
\begin{align*}
    \eta_f^1 &= 1 \\
    \tilde{K}^n &= N_{\eps\eta^n_f}(K^n_x), \ L_f^n = \max\Set{1, \sup_{u \in K_u}{\nu_{\tilde{K}^n}(u)}} \\
    \eta_f^{n + 1} &= 1 + L_f^n\eta_f^n,
\end{align*}
and let
\begin{align*}
    K &= K_x \cup \bigcup_{n = 1}^{N - 1} \tilde{K}^n, \\
    \eps_{n} &= \frac{1}{2^{N - n} \prod_{k=n}^{N - 1}{L_f^{k}}}\eps, \ n = 1,\dots,N.
\end{align*}
Pick a neural network $g \in \Nets^{d_{x} + d_{u}, d_{x}}_{\sigma}$ such that
\begin{equation}\label{eq:dynamics-approx}
    \sup_{x \in K, u \in K_u} \Norm{f(x, u) - g(x, u)} < \min_{n=1,\dots,N}{\eps_n}.
\end{equation}
In particular,
$\sup_{x \in K, u \in K_u}\Norm{f(x, u) - g(x, u)} < {\eps}$.

Now, pick $x \in K_x$ and $u \in \SeqSp{K_u}$.
We have (omitting the $(x, u)$ arguments since they are fixed everywhere)
\begin{align*}
    &\Norm{\dflow_f(n+1, x, u) - \dflow_g(n+1, x, u)} \\
    &\quad= \Norm{f(\dflow_f(n)) - g(\dflow_g(n))} \\
    &\quad\leq \Norm{f(\dflow_f(n)) - f(\dflow_g(n))} + \Norm{f(\dflow_g(n)) - g(\dflow_g(n))}
\end{align*}
Assuming that ${\Norm{\dflow_f(n) - \dflow_g(n)} < \eps\eta_f^n}$,
we have ${\dflow_g(n) \in \tilde{K}^n \subset K}$ and so
\begin{align*}
    \Norm{\dflow_f(n+1) - \dflow_g(n+1)} &< L_f^n\Norm{\dflow_f(n) - \dflow_g(n)} + \eps \\
    &\leq \eps\eta_f^{n + 1}.
\end{align*}
Since~\eqref{eq:dynamics-approx} implies
\begin{equation*}
    \Norm{\dflow_f(1,x,u) - \dflow_g(1,x,u)} < \eps \, (= \eps \eta^1_f),
\end{equation*}
by induction we have that $\Norm{\dflow_f(n,x, u) - \dflow_g(n, x, u)} < \eps\eta_f^n$
for ${n = 1, \dots, N-1}$, so that ${\dflow_g(n, x, u) \in \tilde{K}^n}$.

Now, we show by induction that
\begin{equation*}
    \Norm{\dflow_f(n, x, u) - \dflow_g(n, x, u)} < \eps_n
\end{equation*}
for each $n \geq 0$.
For $n = 1$ this holds by~\eqref{eq:dynamics-approx}:
\begin{equation*}
    \Norm{\dflow_f(1, x, u) - \dflow_g(1, x, u)} =
    \Norm{f(x, u_0) - g(x, u_0)} < \eps_1.
\end{equation*}
Assume that $\Norm{\dflow_f(n, x, u) - \dflow_g(n, x, u)} < \eps_n$.
Then
\begin{align*}
    &\Norm{\dflow_f(n+1, x, u) - \dflow_g(n+1, x, u)} \\
    &\quad\leq \Norm{f(\dflow_f(n)) - f(\dflow_g(n))} + \Norm{f(\dflow_g(n)) - g(\dflow_g(n))} \\
    &\quad< L_f^n \eps_n + \eps_n = \frac{\eps_{n + 1}}{2} + \eps_n \leq \eps_{n + 1}.
\end{align*}
And since $\eps_n \leq \eps$ for $n = 1, \dots, N$, we have that
${\Norm{\dflow_f(n, x, u) - \dflow_g(n, x, u)} < \eps}$, as desired.

Finally, since it is not necessarily the case that $g\in \Nets^0_{\sigma, p}$ for some $p$,
it remains to obtain an equivalent recurrent neural network.
Write $g$ explicitly as
\begin{equation*}
    g(x, u) = T\sigma_{p}(Ax + Bu + b) + c,
\end{equation*}
and rank-factorise $T$ as
\begin{equation*}
    T = M\begin{bmatrix}
        T_1 \\ 0
    \end{bmatrix}
\end{equation*}
with $M \in \Reals^{d_x\times d_x}$ invertible and $T_1 \in \Reals^{r \times p}$ of full row rank.
Then, with
\begin{equation*}
    g_1(x, u) := \begin{bmatrix}
        T_1 \sigma_p(AMx + Bu + b) + c'_1 \\
        c'_2
    \end{bmatrix}, \
    M^{-1}c = \begin{bmatrix}
        c'_1 \\ c'_2
    \end{bmatrix},
\end{equation*}
it follows that
\begin{equation*}
    M\dflow_{g_1}(n, M^{-1}x, u) = \dflow_g(n, x, u)
\end{equation*}
for all $(n, x, u)$.
Let now $T_1^{+}$ be a right inverse of $T_1$ (i.e., $T_1 T_1^{+} = I_r$)
and
\begin{equation*}
    Q := M\begin{bmatrix}
        T_1 & 0 \\ 0 & I_{d_x -r}
    \end{bmatrix}, \ 
    {Q}^+ := \begin{bmatrix}
        T_1^{+} & 0 \\
        0 & I_{d_x - r}
    \end{bmatrix}M^{-1}.
\end{equation*}
Then with
\begin{equation*}
    g_2(z, u) := \begin{bmatrix}
        \sigma_p(AQz + Bu + b) + T_1^+ c'_1 \\
        c'_2
    \end{bmatrix},
\end{equation*}
we get
\begin{equation*}
    Q\dflow_{g_2}(n, Q^{+}x, u) = \dflow_g(n, x, u).
\end{equation*}
Finally, let
\begin{equation*}
    \tilde{A} :=
    \begin{bmatrix}
        AQ \\ 0_{(d_x-r)\times (p + d_x-r)}
    \end{bmatrix}, \
    \tilde{B} :=
    \begin{bmatrix}
        B \\ 0_{(d_x-r)\times d_u}
    \end{bmatrix},
\end{equation*}
\begin{equation*}
    \tilde{b} :=
    \begin{bmatrix}
        b \\ 0_{d_x - r}
    \end{bmatrix}, \
    \tilde{c} :=
    \begin{bmatrix}
        T_1^+ c'_1 \\ c'_2 - \sigma_{d_x-r}(0)
    \end{bmatrix}
\end{equation*}
and define the maps ${\gamma: \Reals^{p + d_x - r} \to \Sp{x}}$ and
${\beta: \Sp{x} \to \Reals^{p + d_x -r}}$
as
\begin{equation*}
    \gamma(z) := Q(z + \tilde{c}), \ \beta(x) := {Q}^{+}x - \tilde{c}.
\end{equation*}
Then with $d_z := p + d_x - r$ and
\begin{equation*}
    h(z, u) := \sigma_{d_z}(\tilde{A}z + \tilde{B}u + \tilde{b} + \tilde{A}\tilde{c})
\end{equation*}
we get
\begin{equation*}
    \gamma(\dflow_{h}(n, \beta(x), u)) = \dflow_g(n, x, u),
\end{equation*}
so that $h \in \Nets^{0}_{\sigma, d_z}$ and
\begin{equation*}
    \Norm{\gamma(\dflow_{h}(n, \beta(x), u)) - \dflow_f(n, x, u)} < \eps
\end{equation*}
for all $x \in K_x$, $u \in S(K_u)$ and $n = 0, \dots, N$, as desired.
Note also that $\gamma$ and $\beta$ have the desired properties.
\end{proof}

\section{Proof of Theorem~\ref{thm:ua-flows}}\label{sec:ua-flows}
We begin with some intuition on the definition of $\Psi$ in~\eqref{eq:psi-def} and the stated assumptions.
Let $\hat\Flow \in \Hyp$ and let $\beta, h, \gamma$ be the corresponding networks
as in~\eqref{eq:hyp}.
Assume for the moment that ${\gamma = \beta = {\Id}_{\Sp{x}}}$.
In the first control period, i.e., for $0 < \tau \leq 1$ it holds that
\begin{align*}
    &\Flow(\tau\Delta, x, u_{\omega}) - \hat\Flow(\tau\Delta, x, u_\omega) \\
    &\quad\quad= \Phi(\tau, x, \omega) - \Bkt{(1 - \tau)x + \tau h(\tau, x, \omega)} \\
    &\quad\quad= \tau{\Par{h(\tau, x, \omega) - \Bkt{x + \tau^{-1}(\Phi(\tau, x, \omega) - x)}}}\\
    &\quad\quad= \tau{\Par{h(\tau, x, \omega) - \Psi(\tau, x, \omega)}}.
\end{align*}
Furthermore, note that $\Phi(1, x, \omega) = \Psi(1, x, \omega)$,
so if we replace $\Phi$ by $\Psi$ in~\eqref{eq:flow-recursive} we get the same result,
provided we interpolate the final state, that is,
\begin{align*}
    \Flow(t, x, u) &= \Phi(\tau_t, x_{k_t}, \omega_{k_t}) =
    (1-\tau_t) x_{k_t} + \tau_t\Psi(\tau_t, x_{k_t}, \omega_{k_t}).
\end{align*}
This motivates the idea that we should approximate the discrete dynamical system
obtained by iterating $\Psi$:
\begin{equation*}
    x_{k + 1} = \Psi(\tau_{k}, x_k, \Param_{k}), \ k \geq 0.
\end{equation*}
Using the recursion map notation, we can equivalently write
\begin{equation*}
    x_{k} = \dflow_\Psi(k, x_0, (\tau_{k},\Param_{k})_{k = 0}^{\infty}).
\end{equation*}
By Theorem~\ref{thm:ua-discrete}, there exists a network ${h \in \Nets_{\sigma, d_z}^0}$ and affine maps $\gamma, \beta$
such that
\begin{equation}\label{eqn:ua-disc-flow}
    \Norm{
        \gamma(
            \dflow_h( n, \beta(x), (\tau_k, \Param_k))) 
            - \dflow_{\Psi}(n, x, (\tau_k, \Param_k))
    } < \eps
\end{equation}
for ${n = 0, \ \dots, \ k_T + 1}$ and any ${x \in K_x}$ and ${(\tau_k, \omega_k) \in \SeqSp{[0, 1]\times K_\Param}}$.
Let $\hat\Flow \in \Hyp$ be defined by these three networks according to~\eqref{eq:hyp},
and recall that $\gamma \circ \beta = \Id_{\Sp{x}}$.

Fix $x \in K_x$, $u \in \Controls(K_\Param)$ and $t \in [0, T]$.
Let $(\Param_k)$ be a sequence parameterising the control $u$ and define
\vspace{-.5em}
\begin{align*}
    z_0 &= \beta(x) \\
    z_{k + 1} &= h(1, z_{k}, \Param_k), \ 0 \leq k < k_t \\
    z_{k_t + 1} &= h(\tau_t, z_{k_t}, \Param_{k_t}).
\end{align*}
Then, as before
\vspace{-.5em}
\begin{equation*}
    z_n = \dflow_{h}(n, \beta(x), (\Tau^t_k, \Param_k)), \, 0 \leq n \leq k_t + 1
\end{equation*}
(recall the definition of $(\Tau^t_k)$ in~\eqref{eq:tau-seq}).
It follows from~\eqref{eqn:ua-disc-flow} that
\vspace{-.5em}
\begin{equation*}
    \Norm{ \Flow(k\Delta, x, u) - \gamma(z_k) } < \eps
\end{equation*}
for $k = 0, \dots, k_t$ and with $x_{k_t} := \Flow(k_t\Delta, x, u)$
\vspace{-.5em}
\begin{equation*}
    \Norm{ \Psi(\tau_t, x_{k_t}, \omega_{k_t}) - \gamma(z_{k_t + 1}) } < \eps.
\end{equation*}

Write
\vspace{-.5em}
\begin{align*}
    &\Flow(t, x, u) - \hat\Flow(t, x, u) \\
    &~= \Flow(t, x, u) - {\gamma}{\Par{(1 - \tau_t)z_{k_t} + \tau_t z_{k_t + 1}}} \\
    &~= \Phi(\tau_t, x_{k_t}, \omega_{k_t}) - {\gamma}{\Par{(1 - \tau_t)z_{k_t} + \tau_t z_{k_t + 1}}} \\
    &~= x_{k_t} + {\tau_t}{\Par{\Psi(\tau_t, x_{k_t}, \omega_{k_t}) - x_{k_t}}} \\
    &\quad\quad{}- {\gamma}{\Par{(1 - \tau_t)z_{k_t} + \tau_t z_{k_t + 1}}} \\
    &~= (1 - \tau_t)x_{k_t} + {\tau_t}{{\Psi(\tau_t, x_{k_t}, \omega_{k_t}) }} \\
    &\quad\quad{}- (1 - \tau_t){\gamma}(z_{k_t}) - \tau_t \gamma(z_{k_t + 1}) \\
    &~= (1 - \tau_t)(x_{k_t} - \gamma(z_{k_t})) + {\tau_t}{\Par{\Psi(\tau_t, x_{k_t}, \omega_{k_t}) - \gamma(z_{k_t + 1})}}.
\end{align*}
If $t < \Delta$ then $k_t = 0$, so that
\vspace{-.5em}
\begin{align*}
    &\Flow(t, x, u) - \hat\Flow(t, x, u) \\
    &\quad= (1 - \tau_t)(x - \gamma(z_{0})) + {\tau_t}{\Par{\Psi(\tau_t, x, \omega_{0}) - \gamma(z_{1})}} \\
    &\quad= (1 - \tau_t)(x - \gamma(\beta(x))) + {\tau_t}{\Par{\Psi(\tau_t, x, \omega_{0}) - \gamma(z_{1})}} \\
    &\quad= {\tau_t}{\Par{\Psi(\tau_t, x, \omega_{0}) - \gamma(z_{1})}},
\end{align*}
and thus
\vspace{-.5em}
\begin{align*}
    &\Norm{ \Flow(t, x, u) - \hat\Flow(t, x, u) } = {\tau_t}{\Norm{\Psi(\tau_t, x, \omega_{0}) - \gamma(z_{0})}} \\
    &\quad< \eps\tau_t \leq \eps.
\end{align*}
For $t \geq \Delta$, we have
\begin{align*}
    &\Norm{ \Flow(t, x, u) - \hat\Flow(t, x, u) } 
    \leq {(1 - \tau_t)}{\Norm{x_{k_t} - \gamma(z_{k_t})}}\\
    &\quad+ {\tau_t}{\Norm{\Psi(\tau_t, x_{k_t}, \omega_{k_t}) - \gamma(z_{k_t + 1})}} \\
    &\quad< \eps,
\end{align*}
and the proof is complete.


\section{Flows of Differential Equations}\label{sec:ua-odes}
In this section, we consider the class of flows~$\Flow$
arising from a controlled Ordinary Differential Equation (ODE)
of the form
\begin{equation}\label{eq:ode}
    \dot{\xi}(t) = f(\xi(t), u(t)), \ \xi(0) = x.
\end{equation}
If the function $f: \States \times \Sp{u} \to \Sp{x}$ is sufficiently regular,
the flow of such a system is well-defined for all
Borel measurable and essentially bounded controls~\cite{Sontag98},
and satisfies the ODE in the following sense:
\begin{equation}\label{eq:flow-int}
    \Flow(t, x, u) = x + \int_{0}^{t}{ f( \Flow(s, x, u), u(s) ) \dd{s} },
    \ t \in \Reals_{\geq 0}.
\end{equation}
In particular, if $f$ is continuous and the control~$u$ is right-continuous
at time $s \geq 0$,
then it holds that
\begin{equation}\label{eq:flow-derivative}
    \frac{\dd}{\dd t}\bigg|_{t = s} \Flow(t, x, u) = f(\Flow(s, x, u), u(s)).
\end{equation}

We now show that the assumptions in Theorem 1 are satisfied in this case under mild conditions on the input parameterisation~$\alpha$ and the right-hand side~$f$ of the ODE.

\begin{lemma}\label{lem:ode}
    Assume that the functions~$f$ in~\eqref{eq:ode}
    in and~$\alpha$ in~\eqref{eq:control-param} satisfy the following conditions
    \begin{enumerate} [I)]
        \item \label{asn:alpha}
            The function $\alpha$ is measurable, and for each
            $\Param \in \Sp{\Param}$ the function
            $\alpha(\Param, \cdot)$ is bounded on $[0, 1]$
            and right-continuous at $t = 0$.
            Furthermore, the family of functions
            $\Set{\alpha(\cdot, t): \Sp{\Param} \to \Sp{u}: t \in [0, 1)}$
            is equicontinuous, i.e.,
            if $\Param_n \to \Param$ then for any $\eps > 0$
            there exists $N \in \PInts$ such that for all $t \in [0, 1)$
            and $n \geq N$ it holds that
                $\Norm{\alpha(\omega_n, t) - \alpha(\omega, t)} < \eps$.
            
        \item \label{asn:f-rhs}
            The function f is continuously differentiable in $(x, u)$,
            and solutions to~\eqref{eq:ode} exist in the sense of~\eqref{eq:flow-int} 
            for $t \in \Reals_{\geq 0}$, for all $x \in \States$
            and all measurable and essentially bounded controls
            $u: \Reals_{\geq 0} \to \Sp{u}$.
        \end{enumerate}
        Then the flow $\Flow$ associated to the ODE~\eqref{eq:ode}
        satisfies the assumptions of Theorem~\ref{thm:ua-flows}.
\end{lemma}
Recalling the definition of $u_{\omega}$ in~Theorem~\ref{thm:ua-flows},
Assumption~\ref{asn:alpha} implies that $u_{\omega}$
is continuous from the right at $t = 0$ and that
${u_{\omega_n} \to u_{\omega}}$ uniformly when ${\omega_n \to \omega}$.

Assumption~\ref{asn:f-rhs} above is sometimes
referred to as \emph{forward completeness} of~\eqref{eq:ode}.
There is no single condition on $f$ that 
can guarantee forward completeness;
examples of possible conditions are discussed
in~\cite{Sontag98, AngeliSontag99}.
It implies the existence of~$\Flow$ 
satisfying~\eqref{eq:flow-int} for all $t \geq 0$,
and that $\Controls$ can be chosen to be the set of all
measurable essentially bounded functions $u: \Reals_{\geq 0} \to \Sp{u}$.

\begin{proof}
First, we show that Assumption~1 in Theorem~\ref{thm:ua-flows} holds. Let $K_\Param \subset \Sp{\Param}$ be a compact set,
and let ${u \in \Controls(K_\Param)}$ be parameterised by the sequence
$(\omega_k)$.
Let ${a_k := \sup_{t \in [0, 1]}\Norm{\alpha(\omega_k, t)}}$.
Suppose $(a_k)$ is unbounded, and pick a subsequence $(a_{k_j})$
such that $\lim_{j \to \infty} a_{k_j} = \infty$.
By compactness, there is a subsequence $(\omega_{k'_{j}})$
of $(\omega_{k_j})$ with $\lim_{j \to \infty}\omega_{k'_{j}} = \omega \in K_\Param$.
Then, by equicontinuity we have that for $j$ large enough
\begin{equation*}
    \Norm{\alpha(\omega_{k'_{j}}, t)} < 1 + \Norm{\alpha(\omega, t)}, \ t \in [0, 1),
\end{equation*}
which implies that $(a_{k'_{j}})$ is bounded, a contradiction.
Hence $a_k$ is bounded, and so $u$ is bounded.
Since $\alpha$ is measurable, so is $u$, and thus $u \in \Controls$, as desired.

We now show that Assumption~2 in Theorem~\ref{thm:ua-flows} holds. Since $\alpha$ is right-continuous at $t=0$, \eqref{eq:flow-derivative} gives
\begin{equation*}
\begin{aligned}
    &\frac{\dd}{\dd \tau}\bigg|_{\tau = 0} \Phi(\tau, x, \Param) =
    \frac{\dd}{\dd \tau}\bigg|_{\tau = 0} \Flow(\Delta\tau, x, u_\Param) \\
    &\quad=  f(\Flow(0, x, u_\Param), u_{\Param}(0))\Delta
    = f(x, \alpha(\omega, 0))\Delta,
\end{aligned}
\end{equation*}
and hence $\Phi$ is differentiable from the right at $\tau = 0$, as desired.


Finally, we show that Assumption~3 in Theorem~\ref{thm:ua-flows} holds. Let $\Psi_0(\tau, x, \omega) = \Psi(\tau, x, \omega) - x$.
We will show that the differential of $\Psi_0$ with respect to $x$ is continuous, 
and thus bounded on compact sets, from which it follows that $\Psi_0$ (and thus $\Psi$)
is locally Lipschitz.
The remainder of the proof requires a few additional properties of the flow $\Flow$ which we state in the following sublemmata.
\begin{sublemma}
Let $(\omega_n) \subset \Sp{\Param}$
and $(x_n) \subset \States$ be such that 
${\omega_n \to \omega}$ and ${x_n \to x \in \States}$.
Then ${\Flow(t, x_n, u_{\omega_n}) \to \Flow(t, x, u_{\omega})}$
uniformly in $t \in [0, \Delta]$.
\end{sublemma}
\begin{proof}
    See Sontag~\cite{Sontag98}, Theorem~1.
\end{proof}
\begin{sublemma}
    The flow $\Flow$ is differentiable with respect to the initial condition $x$ and and its differential with respect to $x$, $D_x \Flow$, satisfies
    \begin{equation*}
        D_x\Flow(t, x, u)\xi = \lambda_{x, u}(t; \xi)
    \end{equation*}
    for all $\xi \in \Reals^n$,
    where $\lambda_{x, u}$ is the solution of the linear boundary value problem
    \begin{equation}\label{eq:variations}
    \begin{aligned}
        \dot{\lambda}_{x, u}(s; \xi) &= D_x{f}(\Flow(s, x, u), u(s))\lambda(s; \xi) \\
        \lambda_{x, u}(0; \xi) &= \xi.
    \end{aligned}
    \end{equation}
    Equivalently, $D_x\Flow(t, x, u) = \Lambda_{x, u}(t)$ where $\Lambda_{x, u}$ is the state transition matrix associated to the linear system~\eqref{eq:variations}.
\end{sublemma}
\begin{proof}
    See Sontag~\cite{Sontag98}, Theorem~1.
\end{proof}
We also need the following result on the continuity of solutions of linear ODEs with respect to the coefficient matrix.
\begin{sublemma}
\label{lma:linear-sensitivity}
    Let $x, z: [t_1, t_2] \to \Reals^{d_x}$ satisfy
    \begin{equation*}
    \begin{aligned}
        \dot{x}(t) &= A(t)x(t)\\
        \dot{z}(t) &= B(t)z(t)
    \end{aligned} \quad\quad t \in (t_1, t_2)
    \end{equation*}
    with $A, B: [t_1, t_2] \to \Reals^{d_x \times d_x}$ measurable and essentially bounded.
    Then, with ${d(t) = x(t) - z(t)}$ and $D(t) := A(t) - B(t)$,
    \begin{align}\label{bound}
        \resizebox{0.49\textwidth}{!}{
        ${\displaystyle\Norm{d(t)} \leq \Par{\Norm{d(t_1)} +
        \int_{t_1}^{t}{\Norm{D(s)}\Norm{x(s)}\dd s}}}
        \mathrm{e}^{\int_{t_1}^{t}{\Norm{B(s)}\dd s}}$
        }
    \end{align}
    for $t \in [t_1, t_2]$.
    
\end{sublemma}
\begin{proof}
    Write
    \begin{align*}
        d(t) = d(t_1) + \int_{t_1}^{t}{\Bkt{B(s)d(s) + (A(s) - B(s))x(s)}\dd s},
    \end{align*}
    so that
    \[
    \resizebox{0.49\textwidth}{!}{
        $\displaystyle\Norm{d(t)} \leq \Norm{d(t_1)} + \int_{t_1}^{t}{\Norm{B(s)}\Norm{d(s)}\dd s} + \int_{t_1}^{t}{\Norm{D(s)}\Norm{x(s)}\dd s},$
    }
    \]
    and Gr\"{o}nwall's inequality gives the desired result.
\end{proof}

We shall use the bound~\eqref{bound} to show that $\Lambda_{x, u_{\omega}}$ is continuous with respect to $(x, \omega)$.
To this end, let $x_n \to x$ and $\omega_n \to \omega$ and, for $t \in [0, \Delta]$, set 
\begin{align}
 A(t) =& D_x f(\Flow(t, x, u_{\omega}), u_{\omega}(t)),\label{eq_A}\\
 B_n(t) =& D_x f(\Flow(t, x_n, u_{\omega_n}), u_{\omega_n}(t)).\label{eq_Bn}
\end{align}
By continuity of $\Flow$ and $\alpha$, there exist compact sets $K'_x$ and $K'_u$ such that
$\Flow(t, x, u_{\omega}), \Flow(t, x_n, u_{\omega_n}) \in K'_x$ and
$u_{\omega}(t), u_{\omega_n}(t) \in K'_u$ holds for all $n$ and $t \in [0, \Delta]$.
Let
\begin{equation} \label{eq_Fbar}
    \bar{F} := \sup{\Set{\Norm{D_x f(z, u)}: \ z \in K'_x, \ u \in K'_u}}.
\end{equation}
We have $\bar{F} < \infty$, by continuity of $D_x f$, and $\Norm{B(s)} \leq \bar{F}$ for $s \in [0, \Delta]$.
Sublemma \ref{lma:linear-sensitivity} applied on the interval $[0, t]$ now gives
\begin{align*}
    &\Norm{\lambda_{x, u_{\omega}}(t; \xi) - \lambda_{x_n, u_{\omega_n}}(t ;\xi)} \\
        &\quad\leq \mathrm{e}^{t \bar{F}}\int_{0}^{t}\Norm{A(s) - B_n(s)}\Norm{\lambda_{x, u_{\omega}}(s; \xi)}\dd s \\
        &\quad\leq \mathrm{e}^{t \bar{F}} \sup_{s \in [0, \Delta]}\Norm{\Lambda_{x, u_{\omega}}(s)} \Par{
             \int_{0}^{t}{\Norm{A(s) - B_n(s)}\dd s}
        }\Norm{\xi},
\end{align*}
so that
\begin{equation}\label{eq:bound-fund-sol}
\begin{aligned}
    &\Norm{\Lambda_{x, u_{\omega}}(t) - \Lambda_{x_n, u_{\omega_n}}(t)} \\
        &\quad\leq \mathrm{e}^{t \bar{F}} \sup_{s \in [0, \Delta]}\Norm{\Lambda_{x, u_{\omega}}(s)}{
             \int_{0}^{t}{\Norm{A(s) - B_n(s)}\dd s}
        }
\end{aligned}
\end{equation}
for each $t$.
By continuity of $D_x f$ and the uniform convergence of $\Flow(t, x_n, u_{\omega_n})$ and $u_{\omega_n}$,
$B_n$ converges uniformly to $A$, and thus $\Lambda_{x_n, u_{\omega_n}} \to \Lambda_{x, u_\omega}$ uniformly on $[0, \Delta]$.


Returning to our original goal, for $\tau > 0$ we have
\begin{align*}
    &D_x{\Psi_0}(\tau, x, \omega)
        = \tau^{-1}(D_x\Phi(\tau, x, \omega) - I) \\
        &\quad = \tau^{-1}{\Par{D_x\Flow(\tau\Delta, x, u_\omega) - I}}
        = \tau^{-1}{\Par{\Lambda_{x, u_{\omega}}(\tau\Delta) - I}}
\end{align*}
Due to the continuity of $\Lambda_{x, u_{\omega}}$, $D_x{\Psi_0}$ is continuous in $(\tau, x, \omega)$ for $\tau > 0$.

For $\tau = 0$ we have
    $D_x{\Psi_0}(0, x, \omega) = \Delta{D_x f(x, u_\omega(0))}$.
If $\tau_n \to 0$ with $\tau_n >0$, $x_n \to x$ and $\omega_n \to \omega$ then
\begin{align*}
    &D_x{\Psi_0}(0, x, \omega) - D_x{\Psi_0}(\tau_n, x_n, \omega) \\
        &\quad= \Delta D_x f(x, u_\omega(0)) - \tau_n^{-1}{\Par{\Lambda_{x_n, u_{\omega_n}}(\tau_n\Delta) - I}} \\
        &\quad= {\Delta D_x f(x, u_\omega(0))} - \tau_n^{-1}{\Par{\Lambda_{x, u_\omega}(\tau_n\Delta) - I}} \\ 
        &\quad\quad\quad{}+ \tau_n^{-1}(\Lambda_{x, u_\omega}(\tau_n\Delta) - \Lambda_{x_n, u_{\omega_n}}(\tau_n\Delta)).
\end{align*}
The first term of the last equality goes to zero, hence we are left to investigate the second term.
With $A, B_n, \bar{F}$ defined in \eqref{eq_A}-\eqref{eq_Fbar}, from~\eqref{eq:bound-fund-sol} we find
\begin{equation*}
\begin{aligned}
    &\Norm{\tau_n^{-1}(\Lambda_{x, u_\omega}(\tau_n\Delta) - \Lambda_{x_n, u_{\omega_n}}(\tau_n\Delta))} \\
        &\quad\leq \mathrm{e}^{\tau_n\Delta \bar{F}} \sup_{t \in [0, \Delta]}\Norm{\Lambda_{x, u_{\omega}}(t)} 
        \frac{1}{\tau_n}\int_{0}^{\tau_n\Delta}{\Norm{A(t) - B_n(t)}\dd t}.
\end{aligned}
\end{equation*}

Pick ${\eps > 0}$, and let $n$ be large enough that ${\Norm{A(t) - B_n(t)} < \eps}$ for ${t \in [0, \Delta]}$, so that
\begin{equation*}
\resizebox{0.49\textwidth}{!}{
    $\displaystyle \tau_n^{-1}\Norm{\Lambda_{x, u_\omega}(\tau_n\Delta) - \Lambda_{x_n, u_{\omega_n}}(\tau_n\Delta)}
        \leq \Par{\Delta \mathrm{e}^{\tau_n\Delta \bar{F}}
            \sup_{t \in [0, \Delta]}\Norm{\Lambda_{x, u_{\omega}}(t)}}\eps$
}
\end{equation*}
and $\tau_n^{-1}(\Lambda_{x, u_\omega}(\tau_n\Delta) - \Lambda_{x_n, u_{\omega_n}}(\tau_n\Delta)) \to 0$ as $n \to \infty$, as desired.

\end{proof}


\vspace{-1em}
\section{Numerical Example}\label{sec:num-example}
We illustrate the use of the proposed architecture
for learning the flow map of a FitzHugh-Nagumo oscillator whose dynamics are given by
\begin{equation}
\label{eq:fhn-dynamics}
\begin{aligned}
    \eta\dot{x}_1(t) &= x_1(t) - x_1(t)^3 - x_2(t)+ u(t) \\
    \eta\gamma\dot{x}_2(t) &= x_1(t)+ a - bx_2(t),
\end{aligned}
\end{equation}
where $\eta = 1/50, \, \gamma = 40, \, a = 0.3, \, b = 1.4$.
The control $u$ is piecewise constant (i.e. ${\alpha(\omega, t) = \omega}$)
with ${\Delta = 0.2}$.

We generate data consisting of $N = 300$~trajectories of~\eqref{eq:fhn-dynamics}
on $t \in [0, 20]$ with $x^n(0) \overset{\text{i.i.d.}}{\sim} N(0, I)$
and each input $u^n$ is parameterised as in~\eqref{eq:control-param}
by a sequence $(\omega_k)$ distributed as follows:
\begin{equation*}
\begin{aligned}
    \omega_{40k} &\overset{\text{i.i.d.}}{\sim}\mathrm{LogNormal}(\mu=\log(0.2), \sigma=0.5),\\
    \omega_{j + 40k} &= \omega_{40k}, \ j = 1, \dots, 39
\end{aligned}
\end{equation*}
for $k \in \PInts$.
In other words, the inputs $u^n$ are square waves with a period of 8~time~units
and the amplitude at each period is sampled from a log-normal distribution.
The dynamics~\eqref{eq:fhn-dynamics} are integrated using a
Backward Differentiation Formula method and for each trajectory $K = 300$ trajectory values
$\xi^n_k = \Flow(t^{n}_k, x^n, u^n) + \eps^n_k$ are sampled,
where $\eps^n_k$ is Gaussian measurement noise with standard deviation equal to 0.05.

\begin{figure}[htbp]
    \centering
    \includegraphics[width=0.99\linewidth]{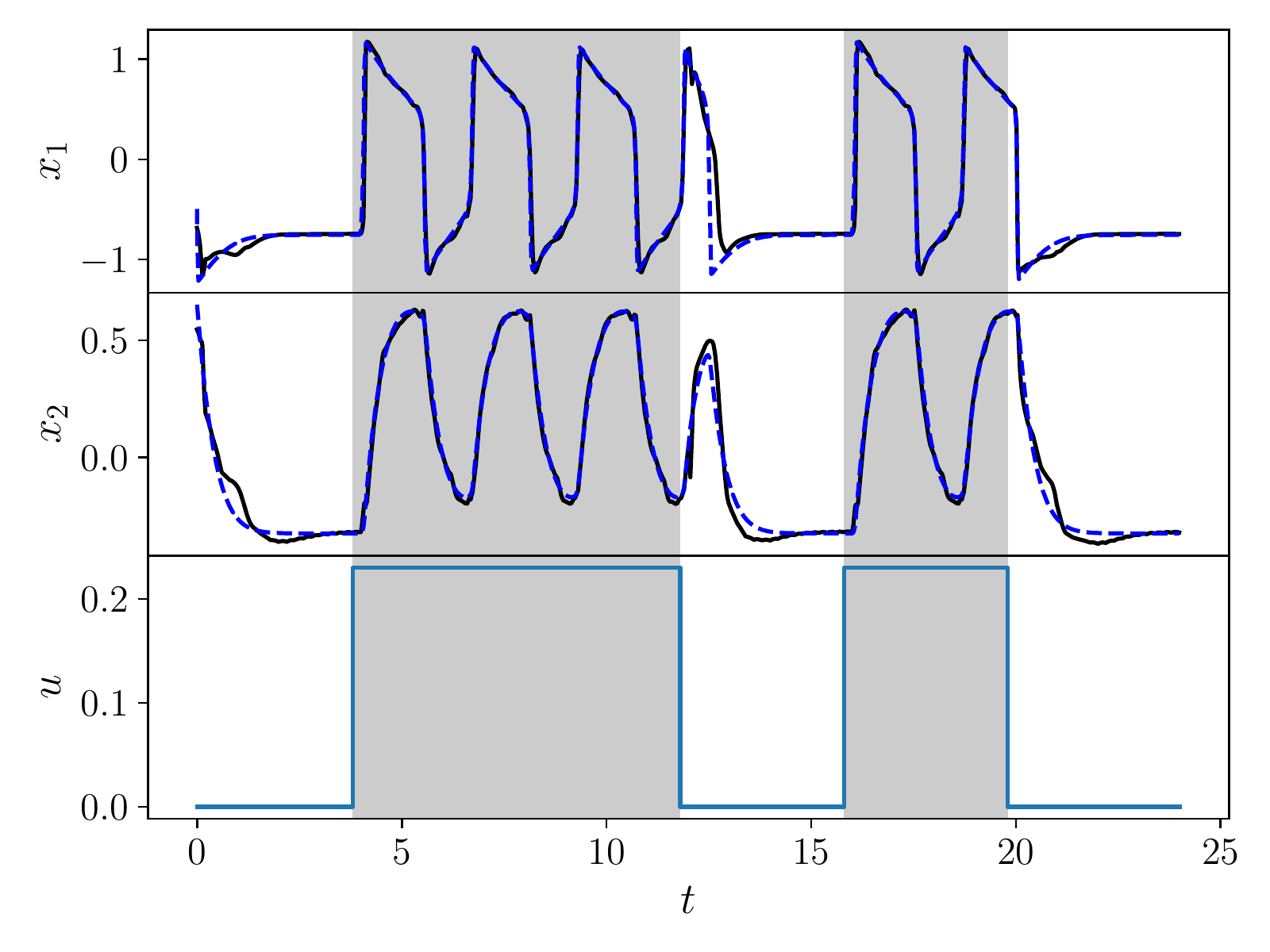}
    \vspace{-2em}
    \caption{
        Real (blue, dashed) and model (black) trajectories of the FitzHugh-Nagumo
        system~\eqref{eq:fhn-dynamics} with the test input $u$.
    }
    \label{fig:fhn-test}
\end{figure}

We train a network defined according to our architecture,
where $h$ is a single-layer long short-term memory recurrent network with 32 hidden states,
and $\gamma, \beta$ are feedforward networks with 3~hidden layers with 32~nodes in
each layer.
The training is done by minimising the mean squared error loss
\begin{equation}\label{eq:mse}
    \frac{1}{N}\sum_{n = 1}^{N}\frac{1}{K}\sum_{k = 1}^K \Norm{
    \xi^n_k - \hat{\Flow}(t^n_{k}, x^n, u^n) }^2
\end{equation}
using the \texttt{Adam} gradient descent algorithm.
Due to space constraints we refer the reader to~\cite{AguiarEtAl23}
for more details on the training procedure.

In Figure~\ref{fig:fhn-test} the real state and predicted state trajectories for
an input $u$ and initial condition not in the training data set are plotted.
The region shaded in grey indicates the times at which the applied input
induces the excitable behaviour of the FitzHugh-Nagumo model~\cite{SepulchreEtAl17},
which is captured by the learned flow model.

\section{Conclusions}\label{sec:conclusion}
We have shown that the RNN-based architecture described in this paper
is a universal approximator of flow functions of dynamical systems with control inputs.
The required assumptions were shown
to hold in the important case of flows of control systems
described by ODEs.
The parameterisation of control inputs,
from which the discrete structure of the flow emerges,
plays a critical role in the architecture.
In effect, our method reflects the fact that continuous-time systems
are most common in practice, 
while the control signals typically arise from discrete-time computation.

A number of avenues for expanding our results are in sight.
We have used one particular way of approximating the discrete-time system
representing the flow function, namely RNNs.
Other sequence models could be applied instead to create variations on our
architecture.
An interesting direction would be to impose stability conditions on the flow
which enables the approximation to hold for unbounded times, as is done
in~\cite{HansonRaginsky20} for continuous-time RNNs.
Estimates on the number of parameters needed to achieve 
a given approximation quality would also be of interest.
Finally, when training learning models in practice the mean squared loss~\eqref{eq:mse} computed on a finite number of trajectory samples
is minimised, which motivates an analysis of the 
sample complexity of learning the flow function.

\vspace{-1em}
\section{Acknowledgments}
This work was supported by the Swedish Research Council Distinguished Professor
Grant~2017-01078 and a Knut and Alice Wallenberg Foundation Wallenberg Scholar Grant.

The computations were enabled by resources provided by the National Academic Infrastructure for Supercomputing in Sweden (NAISS) at C3SE partially funded by the Swedish Research Council through grant agreement no.~2022-06725.

\bibliographystyle{IEEEtran}
\bibliography{IEEEabrv, refs.bib}

\end{document}